\documentclass[aps,12pt,final,notitlepage,oneside,onecolumn,nobibnotes,nofootinbib,
superscriptaddress,noshowpacs,centertags,amsmath,amssymb,amsfonts]{revtex4}
\usepackage[english]{babel}
\newcommand{\pd}{\partial}
\newcommand{\trace}{\mathop{\rm Tr}\nolimits}
\newcommand{\diag}{\mathop{\rm diag}\nolimits}
\newcommand{\beq}{\begin{equation}}
\newcommand{\eeq}{\end{equation}}
\newcommand{\bea}{\begin{eqnarray}}
\newcommand{\eea}{\end{eqnarray}}

\begin{document}
\selectlanguage{english}

\title{Complete integrability of higher-dimensional Einstein
equations with additional symmetry, and rotating black holes.}

\author{\firstname{A.~A.}~\surname{Pomeransky}}

\email{a.a.pomeransky@inp.nsk.su}

\affiliation{Budker Institute of Nuclear Physics, 630090,
Novosibirsk, Russia,\\
and Novosibirsk University}

%\date{\today}

\begin{abstract}
A new derivation of the five-dimensional Myers-Perry black-hole
metric as a 2-soliton solution on a non-flat background is
presented. It is intended to be an illustration of how the
well-known Belinski-Zakharov method can be applied to find
solutions of the Einstein equations in D-dimensional space-time
with D-2 commuting Killing vectors using the complete
integrability of this system. The method appears also to be
promising for the analysis of the uniqueness questions for
higher-dimensional black holes.
\end{abstract}

 \maketitle

\section{Introduction}
There is a number of reasons to be interested in finding exact
black hole solutions of higher-dimensional general relativity. In
string theory our space-time has compactified additional
dimensions. As it was recently noted, the radii of
compactification can be large, and one can check this possibility
experimentally (see \cite{Rubakov:2001} for a review). Another
reason is the recent discovery of a duality between the quantum
gauge theory in usual space-time and the classical gravitation in
five-dimensional anti-de Sitter space-time \cite{Aharony:2000}. At
last, the recent discovery of black rings \cite{Emparan:2001}, the
rotating black hole solutions with the event horizon of unusual
topology $S^1\times S^2$, showed that the no-hair theorems should
be non-trivially generalized in five-dimensional case, and this
generalization can give us better understanding of the reason, why
the no-hair theorems exist in the case of usual four-dimensional
space-time. Motivated by the growing interest in the
higher-dimensional gravitation, we present here a general method
to find the solutions of Einstein equations in the presence of a
sufficient degree of symmetry in all dimensions using their
complete integrability.

The rotating black holes in usual four-dimensional space-time
\cite{Kerr:1963} are described by stationary axisymmetric metrics.
They possess two Killing vectors, corresponding to the time $t$
and the azimuthal angle $\phi$, and the metric can be presented in
the form, independent of $t$ and $\phi$ and dependent only on two
spacial coordinates $\rho$ and $z$. These metrics satisfy the
Einstein equations which in this case have the form of a nonlinear
system for two-dimensional fields. The complete integrability of
this system was shown by Belinski and Zakharov
\cite{Belinski:1978, Belinski:1979}, who found the explicit
formula for N-soliton solutions. The particular cases of these
solutions cover many systems of high physical significance.
Notably, the rotating Kerr black hole corresponds to a 2-soliton
solution \cite{Belinski:1979}. A natural generalization for
arbitrary number of space-time dimensions is to consider
$D$-dimensional space-times with $D-2$ Killing vectors. The
Einstein equations in this case have the same form for all values
of D \cite{Harmark:2004}, and all that is known for $D=4$ (in
particular, the complete integrability and N-soliton solutions)
can be easily generalized to $D>4$. While there exist several
other approaches to the complete integrability in 4D general
relativity (for a recent review and a list of references see
\cite{Kordas:1999}), it is the Belinski-Zakharov method which can
be most easily, in the obvious way, generalized to higher
dimensions. The idea to use the complete integrability in the
higher-dimensional case is not new. Previously, it was applied
most often in the context of Kaluza-Klein theory
\cite{Belinski:1980,Bruckman:1987}. It was applied also to some
particular cases of higher-dimensional theory. These particular
cases include static (Weyl) solutions \cite{Koikawa:2005} and a
class of five-dimensional solutions, which can be reduced to the
four-dimensional case \cite{Mishima:2005}. Closely related models
of gravity with matter fields, arising as low energy limits of
superstring theory, were considered in \cite{Gal'tsov:1995} (see
also references therein).

Among the solutions with the required degree of symmetry there are
rotating black holes in various dimensions. In $(D-1)$-dimensional
space a body (or a black hole) can rotate in
$\left[\frac{D-1}{2}\right]$ mutually orthogonal planes along the
same number of angular coordinates (where square brackets $\left[
\right]$ stand for the integer part). If the black hole is
stationary, the rotation should not change the metric with time.
The metric is thus independent of these
$\left[\frac{D-1}{2}\right]$ angles and each angle corresponds to
a Killing vector. Together with the Killing vector along the time
coordinate this gives $\left[\frac{D+1}{2}\right]$ Killing vectors
for a general rotating black hole in D-dimensional space-time: two
for $D=4$, three for $D=5$ and  $D=6$, and so on. Therefore, the
Einstein equations for black holes in 4 and 5 space-time
dimensions have enough symmetries to be completely integrable,
while for $D\geq 6$ the number of Killing vectors is not
sufficient. The case of $D\geq 6$ corresponds to another setting,
when $D-5$ or $D-4$ of the spacial dimensions are compactified on
circles.

Black holes in five-dimensional space-time are especially
interesting because of the recently discovered black ring
solutions \cite{Emparan:2001}. These solutions, with an unusual
topology of the event horizon, exist in addition to previously
known 5D analogue of the Kerr black hole, the Myers-Perry solution
\cite{Myers:1986}. One may hope, that the complete integrability
by the inverse scattering method can help to answer the arising
questions about uniqueness of regular 5D black hole solutions. The
structure of both the black ring and Myers-Perry solutions
suggests, that they can be interpreted as soliton solutions. As an
illustration of the potential of the method we give here a new
derivation of the $D=5$ Myers-Perry metric as a 2-soliton solution
on a simple static background. While originally, Myers and Perry
found their solution in a certain degree by guessing it, and then
directly verifying its validity, our derivation is based on a
regular method. We were not able to do the same for the general
black ring solution, so this is left for the future work.

The plan of the rest of the paper is as follows. In section II we
remind the reader following the work
\cite{Belinski:1978,Belinski:1979} how the inverse scattering
method can be applied to general relativity in a space-time with
sufficient degree of symmetry, stressing that everything works
equally well for all $D$. In section III we apply this method to
give a new derivation of the 5D Myers-Perry metric as a 2-soliton
solution. We conclude in section IV with a short summary  of the
results and a discussion of future perspectives.

\section{Inverse scattering method and solitons}
In a space-time with $n$ commuting Killing vectors one can always
introduce a coordinate system, in which the metric is independent
of $n$ coordinates. We shall consider a $D$-dimensional space-time
with $D-2$ Killing vectors. In this case we can write down the
metric in the form in which it depends on 2 coordinates only. We
shall denote these coordinates $\rho$ and $z$. If furthermore the
Einstein equations are satisfied, the metric can be written down
in the following simple form \cite{Harmark:2004}:
 \beq\label{metrica}
 -ds^2=g_{ab}dx^a dx^b+f(d\rho^2+dz^2), \;\;\; \det g = -\rho^2,
 \eeq
The Einstein equations for this metric are equivalent to the
following equations for the $D-2\times D-2$ matrix $g_{ab}$:
 \beq\label{system}
 \pd_i (\sqrt{-g}g^{ab}\pd_i g_{bc})=0;
 \eeq
and for the conformal factor $f(\rho,z)$:
 \bea\label{factor}
 \pd_\rho \ln f &=&
 -\rho^{-1}+\frac{\rho}{4}(g_{ab,\rho}g_{cd,\rho}-g_{ab,z}g_{cd,z})g^{ac}g^{bd}
 =-\rho^{-1}+(4\rho)^{-1}\trace(U^2-V^2),\nonumber\\
 \pd_z \ln f &=&
 \frac{\rho}{2}g_{ab,\rho}g_{cd,z}g^{ac}g^{bd}=(2\rho)^{-1}\trace(UV).
 \eea
Here the following notations were introduced for matrices
 \beq\label{uv}
 U=\rho\pd_\rho g\,g^{-1}\,,\;\;\;\; V=\rho\pd_z g\,g^{-1}.
 \eeq
This system (Eqs. (\ref{system}),(\ref{factor})) is well-known for
the usual space-time, $D=4$. For higher dimensions these equations
were derived recently by Harmark \cite{Harmark:2004}. Evidently,
the equations are the same for all dimensions, only the
dimensionality of the matrix $g_{ab}$ depends on D.

The equations (\ref{system}) for $g_{ab}$ do not contain $f$. If
one solves them first, then one can substitute $g_{ab}$ in the
r.h.s. of Eq. (\ref{factor}) and find $f$ solving arising linear
differential equations of first order. The Eqs. (\ref{factor}) for
the conformal factor $f$ are mutually compatible when Eq.
(\ref{system}) is satisfied for the matrix $g_{ab}$.

It is well-known that system (\ref{system}) is completely
integrable. It follows from the fact that Eqs. (\ref{system}) can
be viewed as the compatibility condition for the following
overdetermined system of linear differential equations
\cite{Belinski:1978,Belinski:1979}:
 \begin{gather}
 D_\rho \psi = \frac{\rho U + \lambda V}{\lambda^2
 +\rho^2}\psi\,,\;\;\;\;
 D_z \psi = \frac{\rho V - \lambda U}{\lambda^2 +\rho^2}\psi\,;\nonumber\\
 D_\rho =
 \partial_\rho+\frac{2\lambda\rho}{\lambda^2+\rho^2}\partial_\lambda\,,\;\;\;\;
 D_z =
 \partial_z-\,\frac{2\lambda^2}{\lambda^2+\rho^2}\partial_\lambda\,.
 \label{linear}
 \end{gather}
Here $\psi(\rho,z,\lambda)$ is a complex square matrix, which is
non-degenerate almost everywhere, $U(\rho,z)$ and $V(\rho,z)$ are
real square matrices independent of $\lambda$. The complex
parameter $\lambda$ is called "spectral parameter". It is easy to
check that the above system is compatible if and only if there
exists a matrix field $g(\rho,z)$ (which is identified with
metric) such that $U$ and $V$ can be derived from $g$ via the Eq.
(\ref{uv}), and $g$ satisfies the Eq. (\ref{system}). Note that
the metric $g$ can be easily extracted from $\psi$ as
$g=\psi(\rho,z,0)$.

One can construct new solutions from known solutions by the
following "dressing" procedure. One starts from a known solution
$g_0$ and finds the corresponding $\psi_0$ by solving the linear
equation (\ref{linear}). Then one looks for the new solution
$\psi$ in the form $\psi=\chi \psi_0$. Making this substitution
into the Eq. (\ref{linear}) results in the equations for $\chi$:
  \bea\label{chi}
  D_\rho \chi &=& \frac{\rho U + \lambda V}{\lambda^2 +\rho^2}\chi-\chi\frac{\rho U_0 + \lambda V_0}{\lambda^2
  +\rho^2}\,,\nonumber\\
  D_z \chi &=& \frac{\rho V - \lambda U}{\lambda^2
  +\rho^2}\chi-\chi\frac{\rho V_0 - \lambda U_0}{\lambda^2
  +\rho^2}\,.
  \eea
These equations have the following involution symmetry: if $\chi$
is a solution, then
 $$\chi^\prime=g\widetilde{\chi}^{\,-1}(-\rho^2/\lambda)g_0^{-1}$$
 is also a solution ($\;\widetilde{}\;$ denotes the matrix transposition). In
general, any pair of solutions is related as
$\chi^{\prime}\psi_0=\chi\psi_0 K(w)$, where
$w=(\rho^2/\lambda-\lambda)/2+z$, $D_\rho w=D_z w =0$, but
Belinski and Zakharov demanded an additional condition to be
satisfied: $\chi^\prime=\chi$. This condition can be rewritten as
 \beq
 g = \chi(\lambda)g_0 \widetilde{\chi}(-\rho^2/\lambda)\,,
 \eeq
and it is easy to see, that it guarantees that the matrix $g$ is
symmetric.

To find the solitonic solutions of Eq. (\ref{chi}) one looks for
$\chi$ that are rational functions of the spectral parameter,
making the following ansatz \cite{Belinski:1978,Belinski:1979}:
 \beq\label{ansatz}
 \chi = 1+\sum_k \frac{R_k}{\lambda-\mu_k},
 \eeq
where the positions of poles $\mu_k$ depend on the coordinates
$\rho$ and $z$. Each pole corresponds to a soliton, and the number
of poles is the number of solitons. The coordinate dependence
$\mu_k(\rho,z)$ can be extracted from Eq. (\ref{chi}) by
substituting there Eq. (\ref{ansatz}), noting that the l.h.s. must
have only simple poles in $\lambda$ as the r.h.s. has, and thus
the conditions $D_\rho \mu_k=D_z \mu_k=0$ must be satisfied.
Integrating these differential equations one finds $\mu_k=w_k-z
\pm \sqrt{(w_k-z)^2+\rho^2}$, where the constant $w_k$ is the
soliton position on the $z$ axes. We shall refer to the cases of
signs plus and minus before the square root in this expression as
soliton and antisoliton respectively.

Evidently, $\chi^{-1}$ must be also a rational function of
$\lambda$. For the identity $\chi^{-1}(\lambda)\chi(\lambda)=1$ to
be satisfied at the points $\lambda=\mu_k$,  the matrices
$\chi^{-1}(\mu_k)$ and $R_k$ must be degenerate:
$\chi^{-1}(\mu_k)R_k=0$. This means that $R_k$ factorizes as
 $ R_{k \; a}^{\;\;\;\;b} = n^{(k)}_a m^{(k)b}.$
From Eq. (\ref{chi}) and from the identity
$\chi^{-1}(\lambda)\chi(\lambda)=1$ one can find the vectors
$m^{(k)a}$ and $n^{(k)}_a$. The result is (for more details we
refer the reader to \cite{Belinski:1978,Belinski:1979}):
 \beq\label{mnvectors}
 m^{(k)a} = m^{(k)}_{0b}
 [\psi_0^{-1}(\mu_k,\rho,z)]^{ba}\,,\;\;\;\;
 n^{(k)}_a = \sum_l \mu_l^{-1}D^{kl}N^{(l)}_a\,,
 \eeq
 where the notations
 \beq
 N_a^{(l)} = m^{(l)c} g_{0\,ca}\,,\;\;\;\;
 \Gamma_{kl} = m^{(k)a}g_{0\,ab}m^{(l)b} (\rho^2+\mu_k\mu_l)^{-1}
 \eeq
 were introduced, vectors $m^{(k)}_{0a}$ consist of arbitrary constants,
 and $D^{kl}$ is the inverse of the matrix $\Gamma_{kl}$:
 $D^{km}\Gamma_{ml} = \delta^k_l$. The final expression for the metric is
 \cite{Belinski:1978,Belinski:1979}):
 \beq
 g_{ab}= g_{0\,ab}-\sum_{k,l} D^{kl}\mu_k^{-1}\mu_l^{-1}N^{(k)}_a N^{(l)}_b.
 \eeq
It is useful to write down the expression for the inverse metric
as well:
 \beq
 (g^{-1})^{\,ab}= (g_0^{-1})^{\,ab}-\rho^{-2}\sum_{k,l} m^{(k)a} D^{kl} m^{(l)b}.
 \eeq
 An explicit formula can be written for the conformal factor $f$
 as well \cite{Belinski:1978,Belinski:1979}. It can be shown that
 the ratio $f/f_0$ is proportional to the determinant $\det(\Gamma_{kl})$
 and it depends on the arbitrary constants $m^{(k)}_{0a}$ only through this
 determinant.

In an important particular case, the background metric $g_0$ is
{\it static} (diagonal), and each vector $m^{(k)}_{0a}$ has only
one non-zero component. Then, the resulting metric $g$ is also
diagonal, and it is obtained from $g_0$ simply by multiplying its
diagonal elements $g_{0\,aa}$ corresponding to non-zero elements
of $m^{(k)}_{0a}$ by $-\rho^2/\mu_k^2$.

\section{Myers-Perry black hole in five dimensions}
In the usual four-dimensional space-time the Kerr solution
describing rotating black holes was rederived in
\cite{Belinski:1979} as 2 solitons on the flat Minkowski
background. The Kerr solution has 2 parameters: mass $m$ and
rotation parameter $a$ (the angular momentum is $ma$).
Non-rotating Schwarzschild black hole is a particular case, the
static solution with zero angular momentum $a=0$. While it is very
fortunate, that for Kerr solution the background metric is flat,
there is no any known reason for this to be true {\it a priori},
so this seems to be a mere accident. Indeed, it is easy to see,
that in five-dimensional space-time the analogue of Schwarzschild
solution, the Tangherlini solution \cite{Tangherlini:1963}, cannot
be obtained as a 2-soliton solution on flat Minkowski background.

The metric of Schwarzschild-Tangherlini black hole in
five-dimensional space-time has the form
\cite{Tangherlini:1963,Harmark:2004}:
 \beq\label{schwarzschild}
  g^{Sch}_{\,ab}=\diag \left( -\frac{\mu_{+}}{\mu _{-}}\,,\mu_{-}\,,
  \frac{\rho^2}{\mu_+}\right)\,,\;\;\;\;
  f^{Sch}=\frac{\mu_{-}(\rho^2+\mu_{+}\mu_{-})}{(\rho^2+\mu_{+}^2)(\rho^2+\mu_{-}^2)}\,,
 \eeq
 where $\mu_{\pm}=\sqrt{\rho^2+(z \pm \alpha)^2}-z \mp \alpha\,.$
 It can be obtained as a two-soliton solution on the following
 background metric:
 \beq
  g^{\prime}_{0\,ab}=\diag \left( -\frac{\mu_{+}}{\mu_{-}}\,,
  - \frac{\rho^2}{\mu _-}\,,-{\mu_+}\right).
 \eeq
 This metric is obtained from (\ref{schwarzschild}) by dividing
 the $\phi\phi$-component by $(-\mu^2_{-}/\rho^2)$ and the $\psi\psi$-component by
 $(-\rho^2/\mu^2_{+})$. In this way we effectively remove a
 soliton at $z=-\alpha$ and an antisoliton at $z=\alpha$
 (cf. the end of the previous section). It is convenient to use as the background a simpler
 metric $g_{0\,ab}=-\frac{\mu_{-}}{\mu_{+}}g^{\prime}_{0\,ab}$:
 \beq
 g_{0\,ab}=\diag \left( 1\,, \frac{\rho^2}{\mu_+}\,,\mu_{-}\right),
 \eeq
using the fact, that the multiplication of a background metric by
a function commutes with the operation of putting solitons on this
background. The corresponding solution of Eqs. (\ref{linear}) is
 \beq
 \psi_{0\,ab}=\diag \left( 1\,,\lambda +\frac{\rho^2}{\mu_+}\,,{\mu_-}-\lambda \right).
 \eeq
On this background we place a soliton at $z=-\alpha$ and an
antisoliton at $z=\alpha$. The corresponding vectors $m^{(1,2)a}$
defined in Eq. (\ref{mnvectors}) are:
 \beq\label{mvec}
  m^{(1)a}=(T_{+},0,\frac{\Psi_{+}}{\mu_{-}-\mu_{+}})\,,\;\;\;\;
  m^{(2)a}=(T_{-},\frac{\Phi_{-}\mu_{-}\mu_{+}}{\rho^2(\mu_{-}-\mu_{+})},0)\,,
 \eeq
where $T_{\pm}$, $\Phi_{-}$ and $\Psi_{+}$ are arbitrary
constants. The particular choice $T_{+}=T_{-}=0$ corresponds to
Schwarzschild solution. In Eq. (\ref{mvec}) we have set
$m^{(1)}_{0\phi}=m^{(2)}_{0\psi}=0$. Non-zero values of
$m^{(1)}_{0\phi}$ and $m^{(2)}_{0\psi}$ give a family of singular
solutions with two additional parameters.

It is useful to introduce the prolate spherical coordinates $x$
and $y$, that allows to express $\mu_{\pm},z$ and $\rho$ as {\em
rational} functions of these coordinates \cite{Harmark:2004}:
 \begin{gather}
 \sqrt{\rho^2+(z \pm \alpha)^2}=\alpha (x \pm y)\,,\;z=\alpha x y\,,\\
 \mu_{\pm} = \alpha (x \mp 1)(1-y)\,, \rho^2=\alpha^2(x^2-1)(1-y^2)\,.\nonumber
 \end{gather}

 It is easy to see, that the obtained solution can have
 singularities only on the $\rho=0$ axis. This axis is naturally
 divided into three parts by the positions of solitons at the points $z=\pm \alpha$.
 In the coordinates $(x,y)$ these three parts correspond to the values
 $x=1, y=\pm 1$ \cite{Harmark:2004}.
 These parts can be identified with {\em rods}, introduced in
 \cite{Harmark:2004,Emparan:2001wk}, and the analysis
 of possible singularities on the $\rho=0$ axis can be reduced to the analysis of the
 {\em rod structure} of the solution. The rod structure was defined in \cite{Harmark:2004}
 as follows. Due to the condition $\det g = -\rho^2$, the metric determinant vanishes
 on the $\rho=0$ axis, and the metric has a zero eigenvalue there.
 If there are more  than one zero eigenvalues there would be a curvature singularity.
 It was argued also in \cite{Harmark:2004}, that the corresponding eigenvector can change
 its direction only in a discrete set of points, because otherwise there would be singular
 intervals on the $\rho=0$ axis. These points divide the $\rho=0$ axis in parts
 called rods. The directions of the metric eigenvectors with zero eigenvalues are called
 the {\it directions of the rods}.  The rods get a natural interpretation in terms of solitons.
 The rod endpoints (that were not already present in the background
 solution) coincide with the positions of solitons. At the same time, the existence of an
 explicit relation between the rod directions and the constant vectors
 $m^{(k)}_{0a}$ remains an open question.

 Another question, related to no-hair theorems in five dimensions, arises naturally here.
 Evidently, the general solution of (\ref{system}) is non-solitonic: it corresponds to a
 continuous density of solitons on the $\rho=0$ axis. However, such solutions have singular
 intervals on the axis of symmetry, so they describe not a black hole,
 but the metric outside a matter distribution (a rotating "star").
 The background solution, in its turn, has to be built only from a
 {\it finite} number of rods, otherwise in generic case the resulting metric would be
 singular. Thus, it is natural to suppose that the black holes correspond to
 {\it solitonic} solutions on some simple backgrounds. If this turns out to be true,
 this probably could help to find the most general five-dimensional black hole metric and
 to prove its uniqueness.

 Let us return now to the analysis of the obtained solution. Using the evident freedom in
 rescaling of the arbitrary constants we chose the normalization condition
 $$ \Psi_{+}^2\Phi_{-}^2-16\alpha^2 T_{+}^2 T_{-}^2=1\,.$$
 It is also convenient to introduce the following 3 parameters:
 \beq
 \rho_0^2=4\alpha(4\alpha T_{+}^2+\Psi_{+}^2)(4\alpha
 T_{-}^2+\Phi_{-}^2), \;\;\;
  a_1=4\alpha T_{-}\Psi_{+}, \;\;\; a_2=4\alpha T_{+}\Phi_{-}\,.
 \eeq
 These parameters are not independent:
 $$ \alpha=\frac{1}{4}\sqrt{(\rho_0^2-a_1^2-a_2^2)^2-4a_1^2 a_2^2}\,.$$
 One still has the freedom to make linear transformations in the space of coordinates
 $t,\phi$ and $\psi$. This transformation can be chosen in such a
 way, that the rod structure of the resulting metric matches the
 rod structure of the flat space-time. Namely, we require that the
 rods at $y=1$  and $y=-1$ have directions along the $\phi$
 coordinate and along the $\psi$ coordinate, respectively. The
 resulting linear transformation of the time and angle coordinates
 has the form:
 \bea
 t &=& t_{new}+4\alpha T_{-} \Psi_{+} \phi_{new}+4\alpha T_{+} \Phi_{-}\psi_{new}\,,\nonumber\\
 \phi &=& \Psi_{+}\Phi_{-}\phi_{new}-4\alpha
 T_{-}T_{+}\psi_{new}\,,\;\;\;
 \psi=\Psi_{+}\Phi_{-}\psi_{new}-4\alpha T_{-}T_{+}\phi_{new}\,.
 \eea
 In this way we obtain the metric of the Myers-Perry black hole in 5D space-time:
 \cite{Myers:1986,Harmark:2004}:
 \bea
 g_{00} &=& -(4\alpha x +(a_1^2-a_2^2) y - \rho_0^2)/\omega,\;\;\;
 g_{\varphi \psi}= \frac{1}{2}a_1 a_2 \rho_0^2 (1-y^2)/\omega,\nonumber\\
 g_{0\varphi} &=& -a_1 \rho_0^2 (1-y)/\omega\,,\;\;\; g_{0 \psi} = -a_2 \rho_0^2 (1+y)/\omega,\nonumber\\
 g_{\varphi\varphi}&=& \frac{1-y}{4}(4\alpha x +\rho_0^2+a_1^2-a_2^2+2 a_1^2 \rho_0^2 (1-y)/\omega)\,,\\
 g_{\psi \psi} &=& \frac{1+y}{4}(4\alpha x +\rho_0^2-a_1^2+a_2^2+2 a_2^2 \rho_0^2 (1+y)/\omega)\,,\nonumber
 \eea
where the subscript "$new$" for the coordinates was dropped and
the notation
 \beq
 \omega=4\alpha x +(a_1^2-a_2^2) y + \rho_0^2
 \eeq
was introduced. As it was explained in the previous section, the
conformal factor can be obtained as
 \beq
 f=f^{Sch}\det\Gamma_{kl}/\det\Gamma^{Sch}_{kl}=\frac{\omega}{8\alpha^2
 (x^2-y^2)},
 \eeq
where Schwarzschild solution corresponds to $T_{+}=T_{-}=0$. This
is in agreement with \cite{Harmark:2004,Myers:1986}.

The Myers-Perry solution with a single non-zero angular momentum
($a_2=0$) was rederived recently in \cite{Mishima:2005} using the
complete integrability of the system. In this particular case the
matrix $g_{ab}$ has the block-diagonal form, and the equations
reduce effectively to the four-dimensional case. Then the authors
of \cite{Mishima:2005} applied the results of
\cite{Castejon-Amenedo:1990} to obtain the Myers-Perry metric with
a single angular momentum parameter. In contrast to this previous
work, the present paper considers genuinely five-dimensional case
of a black hole with two non-zero angular momenta, which can not
be reduced to four dimensions.

\section{Conclusions}
The aim of the present paper was to attract attention to the
potential applications of the complete integrability of Einstein
equations in $D$-dimensional space time with $D-2$ commuting
Killing vectors. The integrability can be seen by an obvious
generalization of the well-known Belinski-Zakharov construction
for the usual four-dimensional case. In particular, in this way an
explicit formula for N-soliton solutions can be written down. As
an illustration of the practical usefulness of this method for
finding solutions to the Einstein equations we have derived the
five-dimensional Myers-Perry black hole metric as a 2-soliton
solution on a static background. The method appears promising for
the analysis of the uniqueness of five-dimensional black hole
solutions. It gives a new point of view on the important notion of
the rod structure, identifying the rod end-points with solitons.

The next step would be to find the general black ring metric as a
soliton solution on a simpler background. This is especially
interesting, because the currently known black ring metric, having
only one non-zero angular momentum parameter, appears to be not
the most general one \cite{Emparan:2001}. To this end, one can try
the same approach as the one that was used in this paper to derive
the Myers-Perry solution. One can start from the static black ring
solution, and remove few solitons and antisolitons from it. One
obtains a new static solution, that has to be used as the
background in the dressing procedure. Then one can put the
solitons back on this background at their initial positions, but
this time with generic values of the arbitrary constants
$m^{(k)}_{0a}$. All our attempts in this direction so far did not
result in finding regular black ring solutions, giving only many
singular ones. However, further investigations are needed to see,
whether the regular black ring solutions can be obtained with a
different combination of solitons and background metric or one has
to modify the method.

 \begin{acknowledgments}
 I am grateful to Anatoly Pomeransky and Roman Sen'kov for many useful
 discussions.
 \end{acknowledgments}

\end{document}